\documentclass[aps,prl,amsmath,amssymb,reprint,superscriptaddress,longbibliography]{revtex4-1}

\usepackage{graphicx}
\usepackage{epsf}
\usepackage{bm}
\usepackage{lipsum}
\usepackage{txfonts}
\usepackage[normalem]{ulem}

\newcommand{\bvec}[1]{\mbox{\boldmath $#1$}}
\graphicspath{{./images/}}

\begin{document}

\title{
Local Density Fluctuation Governs Divergence of Viscosity \\ 
underlying Elastic and Hydrodynamic Anomalies in a 2D Glass-Forming Liquid 
}

\author{Hayato Shiba}
\affiliation{
Institute for Materials Research, Tohoku University, Sendai 980-8577, Japan
}

\author{Takeshi Kawasaki}
\affiliation{
Department of Physics,
Nagoya University, Nagoya 464-8602, Japan
}

\author{Kang Kim}
\affiliation{
Division of Chemical Engineering,
Graduate School of Engineering Science, Osaka University, Osaka 560-8531, Japan
}

\date{\today}

\begin{abstract}
If a liquid is cooled rapidly to form a glass, its structural relaxation becomes retarded, producing a drastic increase in viscosity. In two dimensions, strong long-wavelength fluctuations persist, even at low temperature, making it difficult to evaluate the microscopic structural relaxation time. This Letter shows that, in a 2D glass-forming liquid, relative displacement between neighbor particles yields a relaxation time that grows in proportion to the viscosity. In addition to thermal elastic vibrations, hydrodynamic fluctuations are found to affect the long-wavelength dynamics, yielding a logarithmically diverging diffusivity in the long-time limit. 
\end{abstract}

\maketitle
In many two-dimensional ordering phenomena, fluctuations at long wavelengths are so strong that perfect order is destroyed. 
For example, the transition between a liquid and a crystalline solid is continuous or nearly continuous~\cite{Halperin1978,Zahn1999,Han2008,Shiba2009,Kapfer2015,Dullens2017}. Recently, large-scale molecular dynamics (MD) simulations~\cite{Shiba2016,Flenner2015} 
and colloidal experiments~\cite{Illing2017,Vivek2017} have revealed that such long-wavelength fluctuations 
also exist in two-dimensional (2D) liquids that are rapidly cooled toward the glass transition. 
Although retaining a random amorphous structure, elastic vibrations appear as the rigidity emerges with the decrease in temperature.
The excess of low-frequency phonons in two dimensions~\cite{Shiba2016,Mizuno2017} leads to an elastic anomaly,
where the amplitude of thermal vibrations diverge at long wavelengths. Even in the presence of these long-wavelength fluctuations, the microscopic structural relaxations in 2D and 3D supercooled liquids appear to be similar,  
once the effect of these fluctuations has been eliminated by introducing quantities that characterize the local switching between neighbor particles~\cite{Shiba2016,Vivek2017,Illing2017}.

Albeit a similarity of structural relaxation modality between 2D and 3D glass-forming liquids, 
it does not mean that the transport properties, a key to the nature of glass transitions, are similar in between. 
One problem lies in the relationship between the structural relaxation time and the viscosity.
The glass transition is marked by a drastic increase in macroscopic viscosity with decreasing temperature,
which is intimately related to the divergence of the microscopic structural relaxation time. 
As such, theoretical and computational studies have focused on 
the dynamical mechanism of growth in the microscopic structural
relaxation time, most typically the $\alpha$-relaxation
time $\tau_\alpha$~\cite{Stillinger1995,Ediger1996,Harrowell1996,Goetze2008},
defined as the decay time of the
relaxation function for density fluctuations, {\it i.e.}, the intermediate scattering function.
However, in two dimensions, the strong density fluctuation diverges
at long wavelengths and suppresses its intermediate plateau.
Thus, $\tau_\alpha$ no longer represents the microscopic structural relaxation time.
The first problem arises on how to define the relaxation time 
that represents viscous slowdown of the dynamics.

Furthermore, a more intriguing problem lies in the potential role of 
hydrodynamic effects on the transport properties.
For a liquid in two dimensions, a slow $t^{-1}$ decay (the so-called long-time tail) of the velocity and stress autocorrelation functions leads to a {\it hydrodynamic anomaly} that is marked by a logarithmic divergence of transport coefficients such as diffusivity and viscosity~\cite{Ernst1970,Alder1970,Frenkel1991,HansenMcDonald}.
How it alters the transport properties of deeply cooled liquids at the macroscopic level is a fully open issue, and the glass
transition may be influenced by a mechanism that is different from the freezing of
microscopic structural relaxation due to the cage effect.
In fact, it is difficult to distinguish the characteristic time scales
of the long-time tail and the microscopic structural relaxation,
both of which become significantly large upon supercooling.
Such long-wavelength fluctuations derived from hydrodynamics may even possibly prohibit the 2D glass transition.
Therefore, in addition to the effect of elastic fluctuations that inhibit the existence of
2D crystals, it is important to reveal
how macroscopic hydrodynamic fluctuations can influence the microscopic structural
relaxation in 2D glass-forming liquids.

In this Letter, in order to address these issues, we perform  
large-scale MD simulations of a 2D glass-forming liquid to examine how the growth of various relaxation times is related to the divergence of macroscopic viscosity in the presence of long-wavelength fluctuations. The simulations are performed with a particular focus on how the dynamics depend on the system size, so that the anomalous enhancement of elastic and hydrodynamic responses can be characterized.  Simulations are performed for a 2D variant of Kob--Andersen binary Lennard--Jones mixtures~\cite{Kob1995}, in which the composition is 65:35~\cite{Sengupta2013,Flenner2015,Flenner2016,Flenner2019}. The mixture is annealed for a sufficiently long time (maximum of $4\times 10^9$ simulation steps) after rapid cooling to target temperatures in the range $0.4\le T\le 1.0$ in the presence of Langevin heat baths. The production runs are then performed as Newtonian ($NVE$) dynamics simulations to prevent the damping of long-wavelength fluctuations. The data presented in the remainder of this Letter are averaged over four or eight independent simulations~\cite{suppl}. 

\begin{figure}
\includegraphics[width=0.9\linewidth]{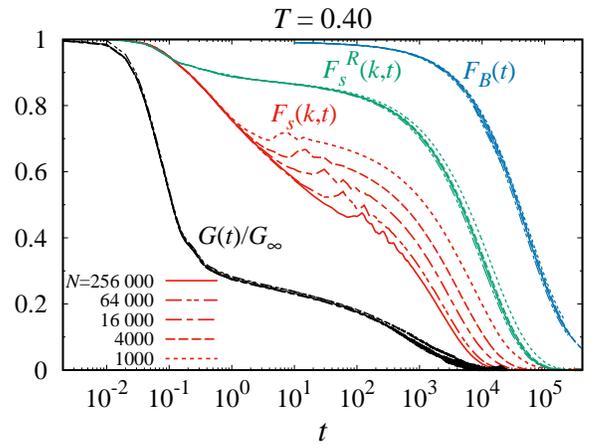}
\caption{ \label{fig:fskt}
  Relaxation functions $F_s(k,t)$, $F_s^{\rm R}(k,t)$, and $F_{\rm B}(t)$
  are plotted for system sizes $N=256000,\ 64000,\ 16000,\ 4000$, and $1000$
  at $T=0.40$. The corresponding stress relaxation function $G(t)$ is
  also plotted (normalized with respect to the instantaneous shear modulus $G_\infty$).
}
\end{figure}

First, we revisit three relaxation functions that have been considered in recent studies~\cite{Flenner2015,Shiba2016,Vivek2017,Illing2017,Flenner2019}. The first is the standard self-intermediate scattering function (SISF)  $F_s(k,t) = (1/N) \langle \sum_{j=1}^N \exp \{ i\bvec{k}\cdot [\bvec{r}_j(t+t_0) - \bvec{r}_j(t_0) ] \} \rangle$, with the wavevector set to $|\bvec{k}| = 2\pi / \sigma_{11}$
so that its decay represents particle movement over a distance of the particle diameter $\sigma_{11}$. However, in two dimensions, long-wavelength elastic vibrations persist, and these enhance the mean-squared thermal displacement $\delta^2$~\cite{Shiba2016,Zhang2019}. 
Figure ~\ref{fig:fskt} shows the SISF for a fixed temperature $T=0.4$, where the
size-dependent behavior is in agreement with previous studies~\cite{Flenner2015,Shiba2016}.
The plateau heights represent the Debye--Waller factor $f_p \sim \exp (-k^2 \delta^2 /2)$~\cite{Pastore1988,suppl},
and tend toward zero as the system size increases. Therefore, the $\alpha$-relaxation time (the decay time of the SISF)
is strongly influenced by long-wavelength fluctuations and cannot represent the microscopic structural relaxation time.

The other two functions are the ``neighbor-relative'' SISF (also known as the ``cage-relative'' SISF~\cite{Illing2017,Vivek2017,Tong2018,Flenner2019,PicaCiamarra2019}) and 
bond relaxation function~\cite{Yamamoto1998a,Shiba2012,Kawasaki2013a,Flenner2016}. These are also plotted in Fig. \ref{fig:fskt}.  The latter 
two functions represent relaxations in the sense that the long-wavelength thermal vibrations are removed. The former is defined by
\begin{equation}
\textstyle F_s^{\rm R}(k,t) = (1/N) \left\langle \sum_{j=1}^N \exp \{ i\bvec{k}\cdot  \Delta\bvec{r}_j^{\rm rel}(t) \} \right\rangle.
\end{equation}
We here introduce the {\it neighbor-relative} displacement
$\Delta\bvec{r}_i^{\rm rel}(t) = (1/N_{\rm n.n.} )  \sum_{j\in n.n.} [\Delta\bvec{r}_i(t) - \Delta\bvec{r}_j(t)]$ (also known as the ``cage-relative'' displacement~\cite{Mazoyer2009,Illing2017,Vivek2017}), where the summation is over $N_{\rm n.n.}$ initially neighboring pairs of particles, indicating the changes in the relative positions.
A similar displacement was considered in previous studies on 2D melting for the same purpose of eliminating 
long-wavelength fluctuations~\cite{Bedanov1985,Zahn2000,Han2008}. 
The bond relaxation function $F_{\rm B}(t)$, 
in contrast, does not involve  displacements of the particles, but simply characterizes the proportion of initially neighboring pairs
that have survived after a certain time~\cite{suppl}.
From the observation that neither function is strongly dependent on the system size, 
contrary to the standard SISF, the effect of long-wavelength fluctuations
is marginal for these relaxation functions, as expected from their definitions.

Along with our aim to relate these microscopic relaxation
functions with the macroscopic viscosity, 
Fig. \ref{fig:fskt} also shows the stress relaxation function (or ``dynamic modulus''~\cite{Flenner2019})
\begin{equation}
G(t) = \frac{V}{k_{\rm B}T}  \langle \sigma_{xy}(t) \sigma_{xy}(0)\rangle,
\end{equation}
where $\sigma_{xy}(t)$ is the off-diagonal stress tensor
(the data are normalized with respect to the instantaneous shear modulus $G_{\infty} = G(0)$).
$G(t)$ exhibits a stretched plateau modulus and no system size dependence at a low temperature of $T=0.4$, as shown in Fig. \ref{fig:fskt}.
In a recent paper, the plateau was found to become unclear, rendering it difficult to evaluate the plateau modulus, for higher temperatures at the onset of slow dynamics, $T\ge 0.7$~\cite{Flenner2019}. 

Next, we define relaxation times and compare them with the transport coefficients  
---the $\alpha$-relaxation time $\tau_\alpha$, neighbor-relative relaxation time $\tau_{\rm R}$, 
and bond relaxation time $\tau_{\rm B}$  can be defined as the decay times of the standard SISF, 
neighbor-relative SISF, and bond relaxation functions~\cite{suppl}. For this purpose, we refer to the Stokes--Einstein (SE) relation $D \eta /T = {\rm const.}$ between the diffusivity $D$ and the viscosity $\eta$. This relation
holds in normal liquids at high temperatures,
but is violated in the deeply supercooled regime~\cite{Tarjus1995,Goree2006,Stillinger2013,Sengupta2013,Kawasaki2013a}.
In simulation studies of glass-forming liquids, because the microscopic structural relaxation time is expected to grow proportionally with the viscosity, 
the left-hand quantity $D\eta /T$ (the so-called SE ratio) is often replaced
by the product of the diffusivity and the relaxation time
$D\tau_\alpha$, and its $T$-dependence is usually examined.
This assumption may break down in two dimensions because 
$\tau_\alpha$ is robustly suppressed by the long-wavelength fluctuations.  
Thus, we calculate the temperature-dependence of the generalized SE ratio $D\tau$
to examine how the three relaxation times change 
with respect to the diffusivity $D$ as the temperature decreases.
We further compare $D\tau$ with the original SE ratio $D\eta /T$
This is done by explicitly calculating the shear viscosity $\eta$ via the Green--Kubo formula  
$\eta = \int G(t)\ dt$~\cite{Sengupta2013,Flenner2019},
for which we require an error estimate because of the slow convergence of this integral~\cite{suppl}.

\begin{figure}
  \includegraphics[width=0.95\linewidth]{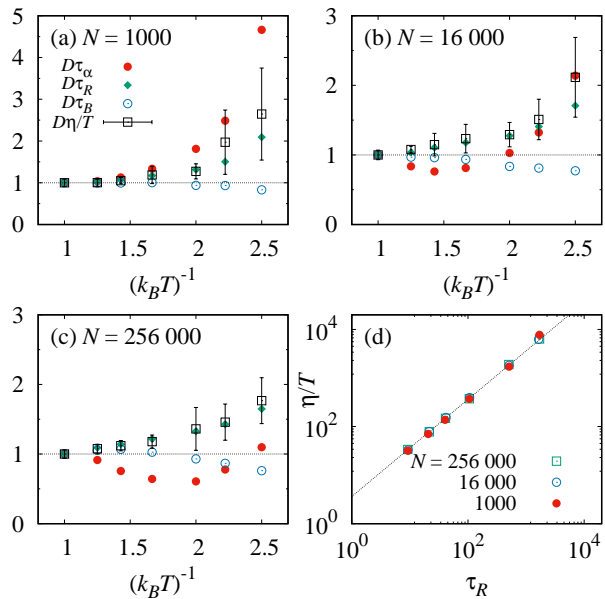}
\caption{ \label{fig:SE}
(a)--(c) SE ratios for (a) $N=1000$, (b) 16000, and (c) 256000 are shown as functions of the inverse temperature.
Error bars indicate the standard deviation arising from the variation in $\eta$ between independent runs.
(d) $\eta /T$ is plotted as a function of the neighbor-relative relaxation time $\tau_{\rm R}$ for different $N$. 
}
\end{figure}

In Fig.\ref{fig:SE} (a)--(c), the original SE ratio $D\eta /T$ is shown
for system sizes of $N=1000$, 16000, and 256000 as functions of the inverse temperature; 
the generalized SE ratios $D\tau_\alpha$, $D\tau_R$, and $D\tau_B$ are also shown.
Although the diffusivity $D$ has a logarithmic dependence on the system size $N$, 
the generalized SE ratios can be meaningfully compared across different values of $N$ because of their similar temperature dependence~\cite{suppl}. 
Firstly, the standard SE ratio increases as the temperature decreases, as in 3D systems.
However, $D\tau_\alpha$ 
exhibits system-size dependence and nonmonotonic temperature dependence
and is clearly decorrelated from {$D\eta /T$}, which is consistent with the results of a previous study~\cite{Sengupta2013}. In contrast, $D\tau_R$ collapses to the standard SE ratio for all system sizes, showing that the neighbor-relative relaxation time $\tau_R$ grows in proportion to the viscosity, satisfying $\tau_R \sim\eta /T$ (see also Fig. \ref{fig:SE}(d)). This relation provides an alternative to $\tau_\alpha \sim\eta /T$~\cite{Tarjus1995,Yamamoto1998b,Stillinger2013,Kawasaki2017},
and thus $\tau_{\rm R}$ clearly takes  the role of the microscopic structural relaxation time. 
We also find that the generalized SE ratio for bond relaxation is preserved, {\it i.e.}, $D\propto \tau_B^{-1}$, indicating that the bond relaxation function is a descriptor for the 2D diffusive motion 
in a similar manner to other 3D supercooled liquids~\cite{Kawasaki2013a,Kawasaki2017}.

\begin{figure}
\includegraphics[width=0.95\linewidth]{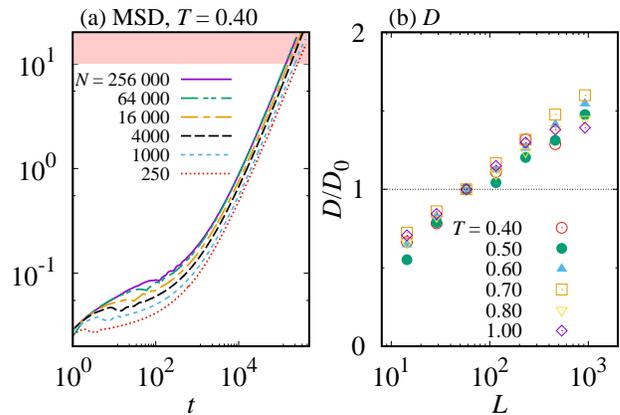}
\caption{ \label{fig:diffusion}
(a) MSDs are plotted for system sizes
  $N=256000,\ 64000,\ 16000,\ 4000,\ 1000,$ and 250 at $T=0.40$.
  Shaded region indicates the fitting region for evaluation of diffusivity.
(b) Diffusivity $D$ as a function of box length $L$.  
   The corresponding particle number ranges over $250 \le N \le 1024000$. 
   For each temperature, the data are normalized with respect to $D_0$, the diffusivity at $N=4000$.
}
\end{figure}

Thus far, we have seen that the diffusivity and viscosity
are linked to time scales associated with local particle motion 
that is irrelevant to the long-wavelength fluctuations.
However, this is not the end of our discussion, and we further
investigate the hydrodynamic effects
on the diffusivity by thoroughly examining the dependence on the system size. 
As shown in Fig. \ref{fig:diffusion}(a) for $T=0.4$, the mean-squared displacements (MSDs) 
$\langle |\Delta \bvec{r}(t)|^2\rangle = (1/N)  \langle\sum_{i=1}^N |\bvec{r}_i(t+t_0) - \bvec{r}_i (t_0)|^2 \rangle$ 
exhibit linear growth and remain dependent on the system size in the long-time limit.
For temperatures $0.4\le T \le 1.0$, we further estimate the diffusivity $D$ by fitting $D = \langle |\Delta \bvec{r}(t)|^2\rangle / 4t$  in the long-time region $10\le\langle |\Delta \bvec{r}(t)|^2\rangle\le 20$ for different system sizes. 
The result is shown as a function of the box length $L$ in Fig. \ref{fig:diffusion}(b).
The diffusivity $D$ grows logarithmically with system size, even at low temperatures.
This size dependence cannot arise from a simple superposition of elastic vibrational fluctuations (the so-called ``Mermin--Wagner fluctuations''~\cite{Illing2017}),  but should be attributed to a different origin.

\begin{figure}
\includegraphics[width=0.8\linewidth]{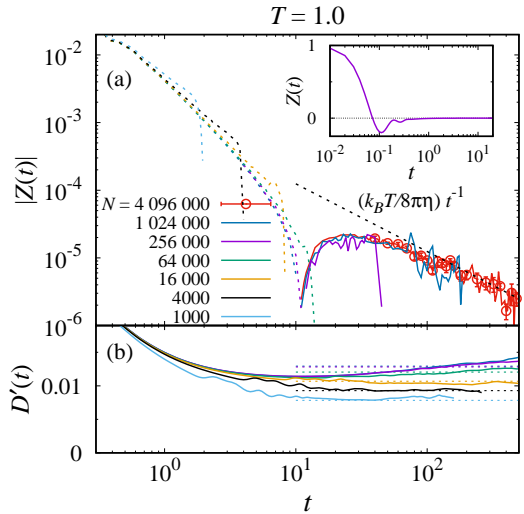}
\caption{ \label{fig:vacf}
(a) Absolute value of VACF $|Z(t)|$ for different system sizes at $T=1.0$.
The solid and dotted lines indicate that the corresponding VACFs assume
positive and negative values, respectively.
The data are only displayed over short time periods for small $N$, for ease of visibility.
The straight line indicates the hydrodynamic long-time tail $Z(t) = (k_{\rm B}T / 8\pi\eta)t^{-1}$.
Inset: Semi-log plot of the raw value of VACF for $N=256000$.
(b) Time-dependent diffusivity, calculated by the Green--Kubo formula $D'(t)
=  \int_0^t Z(t)\ dt$ for different system sizes. 
Dotted lines indicate the diffusivity $D$ evaluated from the MSD in
Fig. \ref{fig:diffusion}(b).
}
\end{figure}

As a statistical measure for the motion of individual particles, 
the velocity autocorrelation function (VACF)
$Z(t) = (1/d) \langle\bvec{v}(t)\cdot \bvec{v}(0) \rangle$~\cite{HansenMcDonald}
contains information regarding the delay in the viscoelastic responses of liquids.
Importantly, VACF is related to the diffusivity via the Green--Kubo formula 
$D = \int_0^\infty Z(t) \ dt$, and should provide clues as to the system-size dependence. 
However, calculating the full resolution of VACF for a glass-forming liquid is a difficult task,
because random motions of the caged particles blur the slow process of diffusion.
In a high-density 3D liquid at a moderately high temperature,
the VACF exhibits a negative correlation in the intermediate time regions~\cite{AlderGass1970,Williams2006,Peng2016},
which can be attributed to the velocity reversal caused by elastic vibrations of tagged particles~\cite{Williams2006}.
Hence, for the present dense 2D liquid, we focus on high temperatures
to demonstrate the crossover from transient elastic response to long-time hydrodynamic decay over a full time range.
Figure~\ref{fig:vacf}(a) shows the VACF for different system sizes $N$ at $T=1.0$.  
Negative correlations exist for all $N$, indicating backward motion originating from elastic
vibrations (see the plot for $N=256000$ in the inset). However, VACF does not simply decay from 
negative values to zero, but becomes positive over a longer time range 
for sufficiently large system sizes, such as $N\ge 256000$. 
By examining a much larger system size ($N=4096000$), 
the long-time limiting behavior is found to be consistent  with the hydrodynamic $t^{-1}$ tail
that appears in normal 2D liquids~\cite{Ernst1970,Alder1970,Frenkel1991,HansenMcDonald,Isobe2008,2017Choi}.
Because the kinematic viscosity is large, {\it i.e.}, $\nu = \eta / (nm) \gg D$, 
the analytical expression can be simplified to $Z(t) = (k_{\rm B}T/8\pi\eta)t^{-1}$.
Both the magnitude and power-law exponent of VACF coincide with this expression.
Therefore, the system size dependence of diffusivity $D$
is attributed to the purely hydrodynamic origin
in the long-time limit where the transient elastic response vanishes. 

Notably, the VACF itself exhibits system-size dependence at times before the $t^{-1}$ power-law tail.  For each $N$, 
the VACF exhibits a systematic decrease before becoming uncorrelated.
In Fig. \ref{fig:vacf}(b), we show the size dependence of the finite-time diffusivity
\begin{equation}
D'(t) = \int_0^t Z(t)\ dt \label{eq:integ}
\end{equation}
in which the long-time limit yields the long-time diffusivity $D = \lim_{t\rightarrow \infty} D'(t)$.
This finite-time diffusivity $D'(t)$ is in good agreement with the diffusivity $D$ evaluated from MSD in the long-time limit. At the same time, it exhibits size dependence in an {\it earlier} time range, before converging to the long-time diffusivity $D$.
Therefore, quite reasonably, the VACF $Z(t)$ itself is affected by the hidden hydrodynamic long-time tail, which has been difficult to find in simulations of 2D glass-forming liquids~\cite{Perera1998,Goree2006,Sengupta2013,Flenner2019}.

The above results suggest that the origin of the logarithmic divergence of diffusivity is the $t^{-1}$ long-time tail 
and that it should exist even at low temperature. The long-time tail 
$Z(t) = (k_{\rm B}T / 8\pi \eta ) t^{-1}$ no longer involves the diffusivity $D$ in its expression
and is free from the self-consistency problem leading to a faster than $t^{-1}$ decay~\cite{Frenkel1991}. 
Therefore, the diffusivity simply diverges as 
\begin{equation}
D\sim \frac{k_{\rm B}T }{ 8\pi \eta } \ln \left(\frac{L}{\sigma_{11}} \right). \label{eq:lnD}
\end{equation}
Hydrodynamic theories also predicts the $t^{-1}$ tail in the shear stress autocorrelation function,
which may cause logarithmic divergence of shear viscosity~\cite{Ernst1970}.
The shear stress autocorrelation functions is calculated for the same temperature  
but still such a power-law tail is not clear with the system size $N=4096000$~\cite{suppl}.

It is worth noting that the analytic expression Eq. (\ref{eq:lnD})
brings about increase of SE ratio  $D\eta / T \sim \ln L$ over all the temperature ranges. 
This ratio increases by an order of magnitude if the system length $L$ 
increases by five orders of magnitude.
While a common expectation seems to be that the effect of long-time tail  is marginal in 
2D glass transitions~\cite{Perera1998},  it is clear that the long-time tail also causes violation of SE relation, 
indicating that it may influence the 2D glass transition.

Finally, we address the relevance of our results to recent studies. 
A recent simulation study on the same 2D Kob--Andersen liquid reported that the mean-square of the neighbor-relative 
displacement ({\it i.e.}, cage-relative MSD) asymptotically approaches the normal MSD in the long-time limit~\cite{Flenner2019}.
Together with our observation of the finite size effects in the MSD,
their result implies that the neighbor-relative displacement is also system-size dependent, finally approaching
the linear behavior of the usual MSD and giving rise to the same diffusivity.
However, the neighbor-relative relaxation time $\tau_{\rm R}$ is short enough such that it
remains virtually unaffected by the long-wavelength fluctuations.
Therefore, our present results fully justify 
the usage of the equivalent cage-relative SISF in recent colloidal experiments~\cite{Illing2017,Vivek2017}.

We note that our results rely on the use of the $NVE$ ensemble to conserve the total momentum. Both the thermal vibrations and the hydrodynamic fluctuations at long wavelengths are suppressed by using the Brownian dynamics~\cite{Flenner2015} or specialized Monte Carlo algorithms~\cite{Berthier2019}. It has long been assumed that glassy dynamics are unaffected by the choice of ensembles, as in 3D liquids~\cite{Gleim1998}, but this is not the case in two dimensions for quantities that are affected by long-wavelength elastic and hydrodynamic fluctuations, including the standard SISF and the MSD.

In conclusion, the dynamics of a 2D glass-forming liquid are covered by hydrodynamic power-law correlations that lead to the logarithmic divergence of diffusivity, in addition to the recently revealed long-wavelength elastic fluctuations arising from the emerging rigidity of the liquid.  Both the elastic and hydrodynamic fluctuations persists at long-wavelengths to produce a concerted effects on the transport properties. Moreover, it is found that the relaxation time defined from the relative displacement between neighbors ($\tau_{\rm R}$) grows in proportion to $\eta /T$ as the temperature decreases,
implying that local density fluctuations govern the drastic increase in viscosity.
The combined elastic and hydrodynamic anomalies are expected to be relevant to
both the existence of the 2D glass transition~\cite{Berthier2019} and the dynamical
features of 2D crystal melting, although further clarification is required in forthcoming studies.

\begin{acknowledgments}
\paragraph{Acknowledgements}
We thank Hajime Tanaka, Kunimasa Miyazaki, and Patrick A. Bonnaud for fruitful discussions.   
This work was supported by JSPS KAKENHI Grant Numbers
JP18K13513 (H. S.), JP19K03767 (T. K.), and JP18H01188 (K. K.). 
H. S. was also financially supported by
Building of Consortia for the Development of Human Resources in Science and
Technology, Ministry of Education, Culture, Sports, and Technology (MEXT), Japan. 
The numerical calculations were performed on the
Cray XC50-LC at the Institute for Materials Research, Tohoku University, Japan, 
the NEC LX406Rh-2 at the Research Center of Computational Science, Okazaki Research Facilities,
National Institutes of Natural Sciences, Japan, and
the SGI ICE XA and HPE SGI 8600 at the Institute for Solid State Physics, University of Tokyo, Japan.
\end{acknowledgments}

\bibliography{2019SKK,2019Suppl}

\end{document}


\title{Supplemental Material for  \\
``Local Density Fluctuation Governs the Divergence of Viscosity Underlying Elastic \\ 
and Hydrodynamic Anomalies in a 2D Glass-Forming Liquid'' }

\author{Hayato Shiba}
\affiliation{Institute for Materials Research, Tohoku University, Sendai 980-8577, Japan}

\author{Takeshi Kawasaki}
\affiliation{Department of Physics, Nagoya University, Nagoya 464-8602, Japan}
\author{Kang Kim}
\affiliation{Division of Chemical Engineering, Graduate School of Engineering Science, Osaka University, Osaka 560-8531, Japan}

\maketitle

\section{Simulation settings and models \label{sec:model} }
Simulations were performed for a 2D variant of Kob-Andersen binary Lennard-Jones mixtures
in which the composition was 65:35~\cite{Sengupta2013,Flenner2015,Flenner2016,Flenner2019,PicaCiamarra2019}. 
The pairwise potential was computed as a function of the interparticle distance $r$:
\begin{equation}
v_{\alpha\beta} (r)=4\epsilon_{\alpha\beta}
\left[ \left( \frac{\sigma_{\alpha\beta}}{r}\right)^{12} - \left(\frac{\sigma_{\alpha\beta}}{r}\right)^6 \right]
\end{equation}
and was truncated and shifted at $r=2.5\sigma_{\alpha\beta}$.
The parameters $\epsilon_{\alpha\beta}$ and $\sigma_{\alpha\beta}$ were set to standard values for arbitrary
combinations of particle species $\{\alpha,\beta\}\in \{1,2\}$~\cite{Kob1995}. 
The number density was fixed at $n =N/L^2 = 1.2$, where $N$ and $L$ denote the total particle number
and the box length, respectively. 
Hereafter, the length, energy, and time are presented in units reduced
by $\sigma_{11},\ \epsilon_{11}/k_B$, and $\sigma_{11}
(m/\epsilon_{11} )^{1/2}$, respectively, and the Boltzmann constant $k_B$ is assumed to be unity.
A liquid at a high temperature ($T=2.5$) was rapidly cooled  
to target temperatures in the range $0.4\le T\le 1.0$.  
The mixture was annealed for a sufficiently long time ($4\times 10^9$ time
steps for the minimum temperature $T=0.4$)
in the presence of a Langevin heat bath. 
Subsequently, production runs were performed.
In this step, we performed Newtonian ($NVE$) dynamics simulations
to avoid the damping of long-wavelength fluctuations.
The simulations were performed for system sizes $250\le N\le 256\ 000$
and temperatures $T=$0.4, 0.45, 0.5, 0.6, 0.7, 0.8, and 1.0,
with a time step of $10^{-3}$.
The results were averaged over eight independent runs
for each $N$ and $T$, except for $T=0.4$, where only four runs were averaged. 
For $N=1\ 024\ 000$ and $4\ 096\ 000$,
additional simulations (four runs for each)
were performed for $0.45\le T \le 1.0$ and $T=1.0$, respectively.

\section{Plateau height of the self-intermediate scattering function}
As a standard relaxation function that is used for characterizing the glassy relaxation,  
the self-intermediate scattering function (SISF) $F_s (k,t)$, with the wavevector $k= 2\pi /\sigma_{11}$,
was calculated (see Fig. 1 of the main text). 
If a plateau appears in the SISF, 
its height corresponds to the so-called Debye-Waller factor~\cite{Pastore1988}
\begin{equation}
f_p = \exp \left(  -\frac{k^2 \delta^2}{d} \right),
\end{equation}
which indicates the strength of the thermal vibrations of a single particle. 
Herein, $\delta^2$ and $d$ represent the mean-squared thermal displacement (MSTD)  and spatial dimensionality, respectively.  
Owing to the Gaussian nature of the thermal fluctuation, MSTD is half its squared amplitude $A^2$, that is, 
$A^2 = 2\delta^2$.
\begin{figure}
\includegraphics[width=0.5\linewidth]{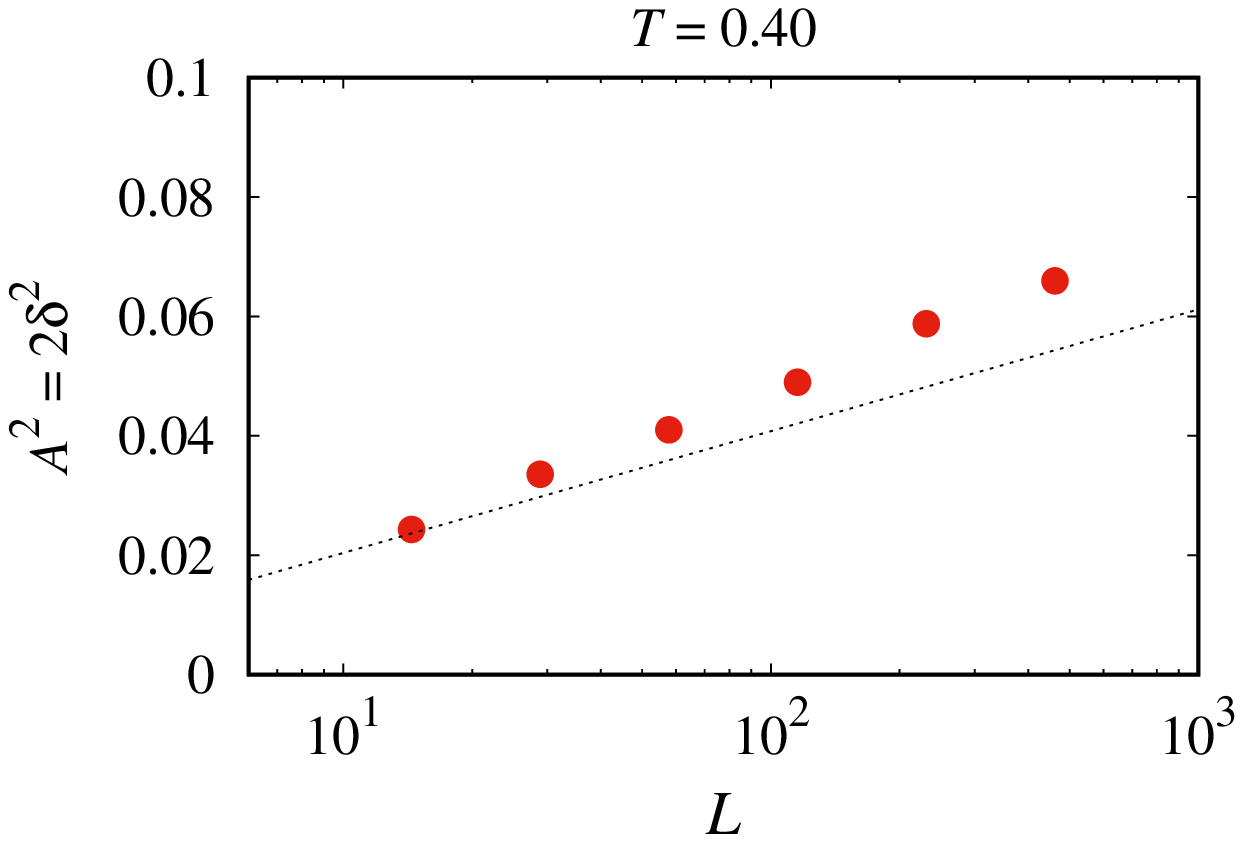}
\caption{\label{fig:DW} 
Squared vibration amplitude $A_p^2  = 2\delta^2$, evaluated from the plateau height of SISF, as a function of the linear box length $L$. 
The broken line is an estimation from the phonon density of states in Eq. (\ref{eq:apsq}).  
}
\end{figure}
In a two-dimensional (2D) glass-forming liquid, the short-time dynamics are dominated by  the solid-like elastic response during supercooling. The MSTD in the plateau region is given by~\cite{Shiba2016}
\begin{equation}
\delta^2 \simeq \frac{k_{B}T}{2\pi  mn}\left( \frac{1}{c_L^2} + \frac{1}{c_T^2}\right) \ln \left( \frac{L}{\sigma_{11}} \right).
\label{eq:apsq}
\end{equation}
This is a consequence of the Debye asymptote giving rise to an abundance of phonon densities of state 
at low frequencies ({\it i.e.}, at long wavelengths). 

For the present 2D Kob-Andersen binary mixture, the longitudinal ($c_L$) and translational 
($c_T$) sound velocities  are estimated to be $c_L = 10.24$ and $c_T = 3.68$. 
In Fig. \ref{fig:DW},  the system-size dependence of the squared vibration amplitude $A^2 = 2\delta ^2$ 
is evaluated from the plateau height of SISF.  Its increase with system size coincides with Eq. (\ref{eq:apsq}),
 which clearly accounts for the origin of the size dependence of SISF. 

\section{Definitions of relaxation times}
In addition to the SISF and the neighbor-relative SISF,
we employ the bond relaxation function~\cite{Yamamoto1998a,Shiba2012,Kawasaki2013a,Shiba2016,Flenner2016},
which characterizes changes in the number of neighbor particles. 
We first count the number of neighbor pairs at the initial time $N_B(t_0)$ satisfying
\begin{equation}
r_{ij}(t_0) = |\bvec{r}_i(t_0) -\bvec{r}_j(t_0) | < 1.2 \sigma_{\alpha\beta}, 
\end{equation}
where $r_{ij} (t_0)$ is the distance between particles $i$ and $j$ at time $t_0$. 
At a later time $t+t_0$,  
we count the number of pairs of particles $N_B (t_0+t; t_0)$ that have become separated: 
\begin{equation}
r_{ij}(t_0+t) = |\bvec{r}_i(t_0+t) -\bvec{r}_j(t_0+t) |> 1.5 \sigma_{\alpha\beta}.
\end{equation}
Using these numbers, the bond relaxation function can be defined as
\begin{equation}
F_B(t) = 1 - \frac{N_B (t_0+t; t_0) }{N_B(t_0)}
\end{equation}

Based on the SISF, neighbor-relative SISF, and the bond relaxation function, 
we evaluated the relaxation times using the following procedures. 
First, the $\alpha$-relaxation time $\tau_\alpha$ was defined as the decay time of standard SISF $F_s(k,t)$.
For the present 2D liquid, the decay in the SISF becomes faster as the system size becomes larger,
thus deviating from the stretched exponential form. 
Therefore, we employ the criterion
\begin{equation}
F_s(k,\tau_\alpha) = 0.1
\end{equation}
for the determination of the $\alpha$-relaxation time $\tau_{\alpha}$. 
The results in the main text do not change appreciably if $F_s (k,\tau_\alpha) = 1/e$ is 
employed, as in a previous study~\cite{Flenner2019}. 
For the neighbor-relative SISF $F_s^R(k,t)$, however, 
the relaxation behavior is well approximated by a double stretched exponential function~\cite{Sengupta2013}
\begin{equation}
F_s^{R}(k,t) = (1-f_c)\exp \left(- \frac{t}{\tau_{s}}\right) + f_c\exp \left[-\left( \frac{t}{\tau_{R}} \right)^\gamma \right],
\end{equation}
where $\tau_{s},\ \tau_{R}$, and $\gamma$ are fitting parameters.
This yields a proper evaluation of the neighbor-relative relaxation time $\tau_{R}$.
By fitting the temperature dependence of $\tau_{R}$ to the 
Vogel—Fulcher-Tamman form $\tau_{R} = B\exp ( A/(T-T_{F}))$, with $B,\ A,$ and $T_{F}$ as the
fitting parameters, the glass transition temperature $T_{F}$ can be estimated to be $0.201(2)$. 

The bond relaxation time $\tau_{B}$~\cite{Yamamoto1998a,Shiba2012}
is evaluated by the usual single stretched exponential fit 
of the bond relaxation function $F_{B}(t)$: 
\begin{equation}
F_{B}(t) =  \exp  \left[ -\left( \frac{t}{\tau_{B}} \right)^{\beta} \right], 
\end{equation}
where $\tau_{B}$ and $\beta$ are fitting parameters. 

\newpage
\section{Stress relaxation function and evaluation procedure of viscosity}
In evaluating the viscosity $\eta$ using the Green-Kubo formula
\begin{equation}
  \eta = \int G(t)\ dt, \quad G(t) = \frac{V}{k_{B}T} \langle \sigma_{xy}(t) \sigma_{xy}(0)\rangle,
\end{equation}   
the stress relaxation function $G(t)$ is calculated from the trajectory
of each independent run and is numerically integrated over time. 
We set the upper limit of the integration to the time at which $G(t)$
becomes less than zero for the first time (in each run).
We averaged the resulting values of $\eta$ 
over eight (or four for $T=0.4$) independent simulation runs,
and estimated the standard deviation.

\begin{figure}[h]
\includegraphics[width=0.55\linewidth]{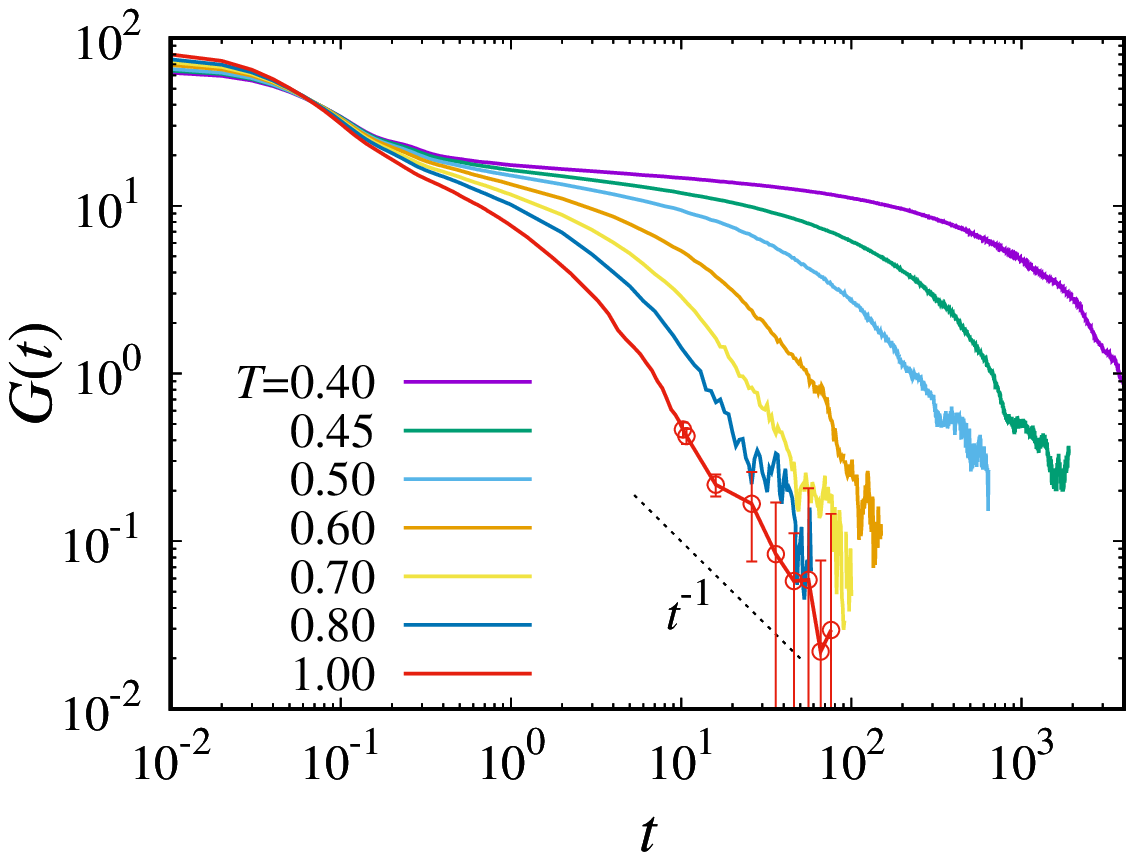}
\caption{\label{fig:sscorr} 
Stress relaxation function $G(t)$ for various temperatures.
The system size is $N=256\ 000$, except for $N=4\ 096\ 000$ at $T=1.0$. 
The errors for $T=1.0$ are evaluated from four independent simulations. 
}
\end{figure}

In Fig. \ref{fig:sscorr}, a logarithmic plot of the stress relaxation functions $G(t)$ 
is given for various temperatures.  The system size is $N=4\ 096\ 000$ for $T=1.0$,
and $N=256\ 000$ for other temperatures. 
We find a formation of plateau, but as noted in Ref. \onlinecite{Flenner2019}, 
it cannot be well fitted by the stretched exponential function for $T\ge 0.7$. 
In a 2D liquid, the so-called hydrodynamic long-time tail 
should also appear in $G(t)$.  
At $T=1.0$, there is a symptom of the power-law tail, but the error is 
still too large to conclude its existence. Therefore, it is left for future large-scale simulations.

However, the variance of $\eta$ mainly arises from the plateau-like regions of $G(t)$,
rather than from the longer-time relaxation close to the cutoff where such behavior might exist.
The viscosity (in terms of the Green-Kubo formula)
does not vary significantly with the long-time tail of $G(t)$, even if it exists,
because the majority of the contribution and statistical errors arise from the intermediate time window. \\ 

\newpage
\section{Temperature dependence of the diffusivity}
\begin{figure}[h]
\includegraphics[width=0.4\linewidth]{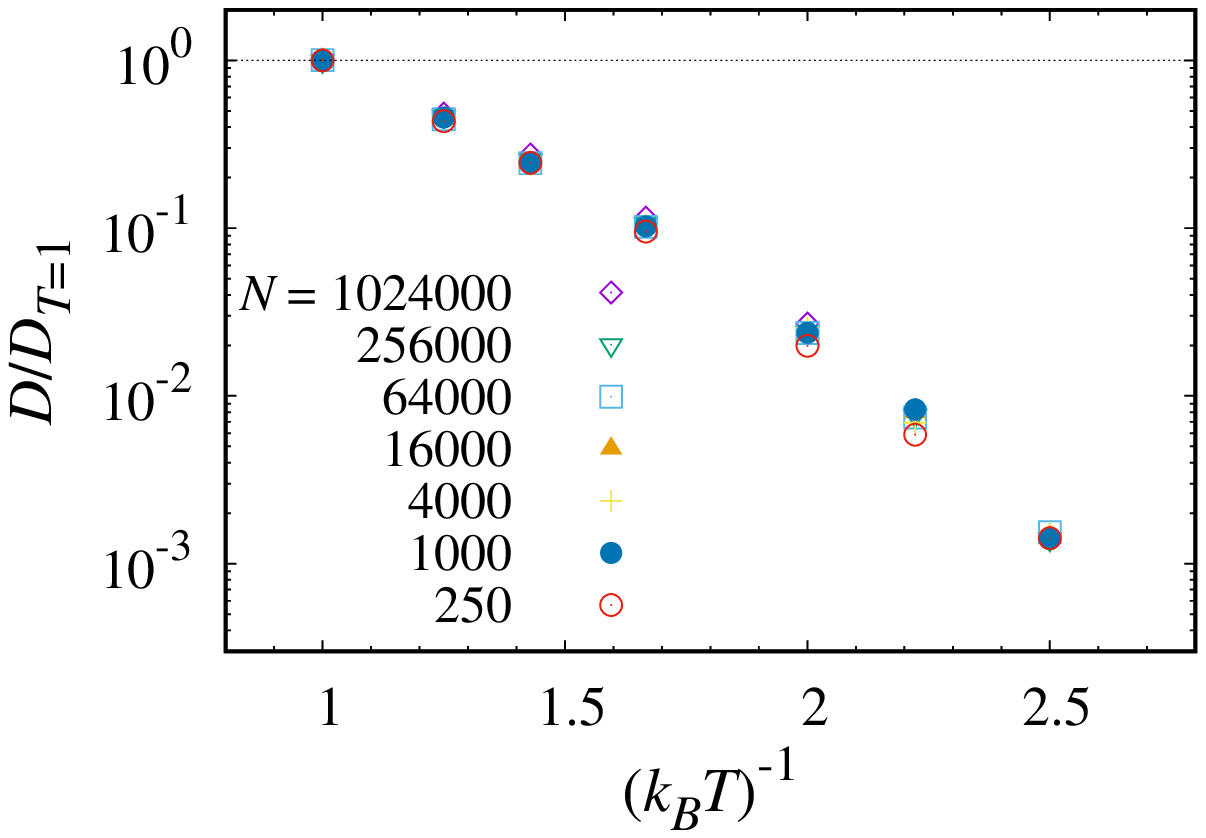}
\caption{\label{fig:DT} 
Diffusivity $D$ evaluated from the MSDs as functions of the inverse temperature for system sizes $250\le N\le 1\ 0240\ 00$.}
\end{figure}

In the 2D glass-forming liquid under consideration, diffusivity $D$ is found to be
influenced by long-wavelength hydrodynamic fluctuations 
that are dependent on the system size, contrary to the behavior in higher dimensions.
Although this situation is beyond what is assumed in the Stokes-Einstein (SE) relation,
the temperature dependence of diffusivity itself remains similar at different system sizes
as the temperature decreases, as shown in Fig. \ref{fig:DT}.
Therefore, the SE ratios ($D\eta /T$ or $D\tau$) can still be employed
to examine the relationship between the diffusivity and the viscosity. \\

\bibliography{2019SKK}